\documentclass[aps,onecolumn,amssymb]{revtex4}
\usepackage{epsfig,amssymb,amsmath,graphicx}

\newcommand{\be}{\begin{equation}}
\newcommand{\ee}{\end{equation}}
\begin{document}
\title{Correlated percolation and tricriticality}
\author{L. Cao$^1$ and J. M. Schwarz$^1$}
\affiliation{Physics Department, Syracuse University, Syracuse, NY 13244}
\date{\today}
\begin{abstract}
The recent proliferation of correlated percolation models---models
where the addition of edges/vertices is no longer independent of
other edges/vertices---has been motivated by the quest to find
discontinuous percolation transitions. The leader in this
proliferation is what is known as explosive percolation. A recent
proof demonstrates that a large class of explosive
percolation-type models does not, in fact, exhibit a discontinuous
transition[O. Riordan and L. Warnke, Science, {\bf 333}, 322
(2011)]. We, on the other hand, discuss several correlated
percolation models, the $k$-core model on random graphs, and the
spiral and counter-balance models in two-dimensions, all
exhibiting discontinuous transitions in an effort to identify the
needed ingredients for such a transition. We then construct
mixtures of these models to interpolate between a continuous
transition and a discontinuous transition to search for a
tricritical point. Using a powerful rate equation approach, we
demonstrate that a mixture of $k=2$-core and $k=3$-core vertices
on the random graph exhibits a tricritical point. However, for a
mixture of $k$-core and counter-balance vertices, heuristic
arguments and numerics suggest that there is a line of continuous
transitions as the fraction of counter-balance vertices is
increased from zero with the line ending at a discontinuous
transition only when all vertices are counter-balance. Our results
may have potential implications for glassy systems and a recent
experiment on shearing a system of frictional particles to induce
what is known as jamming.
\end{abstract}
\maketitle

\section{Introduction}

Percolation is the study of connected structures in disordered networks~\cite{percolation1}. For example, two edges meeting at a vertex form a connected structure called a cluster of size two. As more edges are randomly and independently added to the network, the average cluster size grows until there ultimately exists a spanning cluster in finite-dimensional lattices (or a giant component in random graphs). The onset of a spanning cluster, which is indicative of a transition from a non-spanning to spanning phase, exhibits properties of a continuous phase transition. The simplicity of this nontrivial model allows one to catalog many of its properties such that it is the Ising model of geometrically-driven phase transitions~\cite{percolation2,percolation3}.

While the simplicity of percolation is part of its power, there
has been a recent renaissance in developing models beyond ordinary
percolation in an effort to discover new types of transitions such
as a discontinuous one.  The main driving force behind this
endeavor is what is known as explosive percolation~\cite{explosive
science,Hans,explosive3,explosive4,explosive5}. To
be specific, two edges are considered at random and the edge that
minimizes, for example, the product of the two clusters it joins is then
retained and the other discarded, i.e. a choice has been made as
to which edge to retain. Initial numerical data for this
particular model on random graphs suggested that the percolation
transition is discontinuous such that the emergence of the giant
component is explosive, hence the term explosive percolation.
Because explosive percolation goes beyond ordinary percolation
where there is no choice between edges, a different type of
transition is not necessarily surprising.  However, a recent
mathematical proof shows that the transition in this class of
models involving choice on random graphs is, in fact,
continuous~\cite{mathematicalexplosive,network2}.

Other explosive percolation-type models are being investigated for
the possibility of a discontinuous transition. For instance,
simulations of explosive percolation on finite-dimensional lattices
show signs of a discontinuous phase transition~\cite{Ziff,Ziff2}. In
addition, in light of the recent proof that models involving a
choice (Achlitopas processes) on random graphs exhibit a
continuous transition, some researchers have begun to study other
models beyond ordinary percolation where there exist various
constraints on the occupation of edges (and/or vertices). For
example, the Bohman-Frieze-Wormald model allows for the addition
of an edge provided it participates in a cluster smaller than some
prescribed size with the prescribed size being updated as edges
are added~\cite{Bohman-Frieze-Wormald}. Recent work suggests a
discontinuous transition on random graphs for this
model~\cite{Chen&D'Souza1,chen&D'souza2} and even more recent work
suggests a similar result on finite-dimensional graphs such as the
cubic lattice~\cite{BohmanFriezeWormaldD'souza}.

While work progresses on these more complicated percolation
models, perhaps the simplest beyond ordinary percolation model is
known as $k$-core percolation~\cite{CLR,network1,Jen}. $k$-core percolation
is defined as the following: Each vertex on a graph needs at least
k occupied edges; if the constraint is not obeyed, the vertex and
the edges attached to it are recursively removed until a stable
$k$-core configuration is reached (with every vertex obeying the
$k$-core constraint). It turns out that $k$-core percolation on
random graphs exhibits a continuous transition for $k\leq 2$ and a
discontinuous transition for $k\geq 3$. However, unlike a typical
discontinuous transition there exists not one, but several,
diverging lengthscales such that the transition is an unusual one.

So there indeed exists discontinuous percolation transitions in
mean-field.  What about finite-dimensions? It turns out that
$k$-core percolation on various finite-dimensional lattices, such as
the triangular lattice~\cite{triangular lattice}, either falls
into the same universality class as ordinary percolation or does
not exhibit a transition (but interesting finite-size effects), i.e. $p_c=1$~\cite{pc=1,dawson}, where $p_c$
denotes the critical occupation probability.  Of course, in high
enough dimensions it is conjectured that the $k\geq 3$-core
transition becomes discontinuous~\cite{jen&brooks}. While
obtaining a discontinuous transition with $k$-core in
finite-dimensions has been difficult, a new class of correlated
percolation models, dubbed jamming percolation, with constraints
more complex than $k$-core, has been recently shown to exhibit a
discontinuous transition in
finite-dimensions~\cite{spiralmodel,commentonspiralmodel,replytocommentonspiralmodel,spiralmodel2}.
In addition to exhibiting a discontinuous transition, these models
also exhibit diverging lengthscales that grow either as a power
law or faster than power law such that they, too, are not the
garden variety discontinuous transition.

Since there exists percolation models beyond ordinary percolation
exhibiting a discontinuous transition, it is possible to construct
hybrid models where some fraction of the edges are occupied via
ordinary percolation and the remaining edges are occupied via a
choice method or some other constrained method. Since ordinary
percolation dominates in one limit and the correlated percolation
model, which includes explosive percolation, dominates in the
other limit, it may be possible to locate a tricritical point
bordering the continuous and discontinuous regimes. The proof that
explosive percolation is continuous on random graphs removes the
possibility of a tricritical point on random graphs using a hybrid
of ordinary percolation and explosive percolation. However, the
potential for tricriticality in other correlated percolation
models is intriguing and remains a possibility. For example,
tricriticality has been explored in hybrids of ordinary
percolation and explosive percolation in
finite-dimensions~\cite{Ziff'stricrit}.

Here, we explore the possibility of tricriticality in (1) a
mixture of $k=2$ and $k=3$-core edges on random graphs and (2) a
mixture of $k=3$-core edges and jamming percolation edges in
two-dimensions. Cellai and collaborators have studied the mixture
of $k=2$ and $k=3$-core on locally tree-like graphs and random graphs~\cite{Cellai}. Therefore, we expect agreement between our results obtained using a dynamic rate equation method and theirs~\cite{Cellai}. As for the second case,
jamming percolation models are newer so that we will address some
of their properties and argue why the search for a tricritical
point in such hybrid models may reach a dead end.

While explosive percolation potentially applies to community
networks~\cite{app1} and human protein homology networks~\cite{app2}, $k$-core percolation
may apply to glassy and jamming systems.  The Fredrickson-Andersen
model~\cite{Fredrikson-Andersen} is a kinetically-constrained
model mimicking the caging effect~\cite{caging effect experiment}
in glassy dynamics. The onset of a spanning cluster in $k$-core
percolation corresponds to the onset of a glass transition in the
Fredrickson-Anderson model~\cite{sbt}. Recently, the Fredrickson-Andersen
model was extended to incorporate inhomogeneity in the caging
dynamics, i.e. some particles are able to break out of their cage
if there are less than $k=2$ particles surrounding it, while
others require $k=3$~\cite{Sellitto}. This system exhibits a tricritical point in
mean-field as the average value of $k$ is tuned from two to three,
the properties of which were investigated by Cellai and
collaborators~\cite{Cellai}. As for another application of
$k$-core percolation, the $k$-core condition encodes the scalar
aspect of the principle of local mechanical stability present in
jammed packing. It turns out that the exponents associated with
the mean-field $k\geq 3$-core transition are the same exponents as
measured in the jamming transition~\cite{Jen}.

The paper is organized as follows. In the next section (Section II), we present our results for a mixture of $k=2$ and $k=3$-core on the random graph.  In Section III we address a mixture of $k=3$-core and jamming percolation models in two-dimensions. In Section VI we discuss the implications of our results.

\section{Hybrid $k$-core on random graphs}
\subsection{Revisiting the rate equation approach}
In 1996, Pittel, Spencer, and Wormald~\cite{Wormald3authors} proved that the $k\geq
3$-core transition on random graphs is
discontinuous using rate equations for the $k$-core culling procedure~\cite{Wormald}. This method allows one to easily obtain the
discontinuity and we shall review it here as has been done in Ref.~\cite{Hartmann}.

Consider a random graph with $N$ vertices.  At each time step, one vertex whose edges is less than $k$ is removed from the graph. The notion of time is given by time step, $t=T/N$, where $T$ is the number of algorithmic steps taken so far. The change in time is given by $\Delta t=1/N$, which becomes continuous when $N\rightarrow \infty$. At
time $t$, the number of vertices is $N(T)=(1-t)N$, and the
corresponding distribution of connectivity is $P_z(t)=N_z(T)/N(T)$.

One can
write down the expected change for $N_z$ in the $(T+1)$th step. It
contains two different contributions: 1) The first contribution
corresponds to the removed vertex itself. It appears only in the
equations for $N_z$ with $z<k$. 2) The second contribution is from
the neighbors of the removed vertex. The number of its neighbors
with connectivity $z=k$ will be decreased by 1, hence the total number
of vertices with connectivity $k$ is decreased by 1. The number of its
neighbors with connectivity $z=k+1$ will be decreased by 1, too, leading
to the total number of vertices with connectivity $k$ increased by 1. Therefore, the equation for the change for $N_z$ for all $z$ is: \be
N_z(T+1)-N_z(T)=-\frac{\chi_zP_z(t)}{\overline{\chi}}+\frac{\overline{z\chi}}{\overline{\chi}}[-\frac{zP_z(t)}{Z(t)}+\frac{(z+1)P_{z+1}(t)}{Z(t)}].
\label{eq:master equation} \ee
Here, ${\chi}_z$ equals
one if $z<k$ and zero otherwise. In addition, $\overline{\chi}=\sum_z{\chi}_zP_z(t)$
and $\overline{z\chi}=\sum_zz\chi_zP_z(t)$. The average connectivity is defined as
$Z(t)\equiv\overline{z}=\sum_zzP_z(t)$.

In the thermodynamic limit, the difference equations become differential equations, or
\be
\frac{d}{dt}\{(1-t)P_z(t)\}=-\frac{\chi_zP_z(t)}{\overline{\chi}}+\frac{\overline{z\chi}}{\overline{\chi}}[-\frac{zP_z(t)}{Z(t)}+\frac{(z+1)P_{z+1}(t)}{Z(t)}].
\label{derivative form of master equation} \ee
This infinite set of differential equation is
not easy to solve. However, one can assume that as
vertices get removed, those who have never been touched,
i.e. whose connectivity is greater than $k$, obey Poisson statistics
with an effective connectivity ${\beta}(t)$ and initial
connectivity ${\beta}(0)=c$. In other words,

\be(1-t)P_z(t)=\frac{N_z(T)}{N}=e^{-\beta(t)}\frac{\beta(t)^z}{z!}\,\,\,\,\,\,\,\,
 {\forall}z{\geq}k.\label{Poisson Ansatz}\ee

With this ansatz, one can write the normalization condition as:

\begin{eqnarray} 1&=&\sum_{z=0}^{\infty}P_z(t)\nonumber\\
&=&\sum_{z=0}^{\infty}\chi_zp_z(t)+\sum_{z=k}^{\infty}\frac{1}{1-t}e^{-\beta(t)}\frac{\beta(t)^z}{z!}\nonumber\\
&=&\bar{\chi}+\frac{1}{1-t} F_k(\beta(t))
,\label{eq:Normalization}\end{eqnarray} where \be
F_k(\beta)=1-\sum_{z=0}^{k-1}e^{-\beta(t)}\frac{\beta(t)^z}{z!}
.\ee Moreover, the average connectivity, $Z(t)$, can be written as:
\begin{eqnarray}  Z(t)&=&\sum_{z=0}^{\infty}zP_z(t)\nonumber\\
&=&\sum_{z=0}^{\infty}z\chi_zp_z(t)+\sum_{z=k}^{\infty}\frac{z}{1-t}e^{-\beta(t)}\frac{\beta(t)^z}{z!}\nonumber\\
&=&\overline{z\chi}+\frac{\beta(t)}{1-t} F_{k-1}(\beta(t)).
\label{eq:averageconnectivity}\end{eqnarray}
Finally, the rate equation becomes,
\be
\dot{\beta}(t)=-\frac{\beta(t)}{(1-t)Z(t)}\frac{\overline{z\chi}}{\bar{\chi}}.
\ee
One can also obtain
\be
(1-t)Z(t)=\frac{\beta^2(t)}{c}
\ee
by comparing $\dot{\beta}(t)$ and $\frac{d}{dt}((1-t)Z(t))$ with the latter obtained using the rate equation.

Now, the culling procedure ends when the graph reaches a stable $k$-core configuration. In other words, when
$\overline{\chi}=\overline{z\chi}=0$ for all $z{\geq}k$. Therefore, Eq. 5 and Eq. 6 simplify to
\begin{eqnarray}  1-t_f&=&F_k(\beta_f)\nonumber\\
\frac{\beta_f}{c}&=&F_{k-1}(\beta_f),\label{eq:halting
time}\end{eqnarray}
where $\beta_f=\beta(t_f)$ and we have used Eq. 8.
To study the nature of the transition, for a given $c$ (or initial occupation probability $c=p/N$), one solves for $\beta_f$ using the second equation and then computes $1-t_f$, which yields the fraction of vertices left and is of the order of the giant component should it exist. For $k=2$, the critical initial concentration signalling the onset of the giant component is $c_q=1$. For $c=c_g+\epsilon$ with $\epsilon<<1$, $1-t_f=2\epsilon^2 + \mathcal{O}(\epsilon^3)$, i.e. the transition is continuous.  For $k=3$, $1-t_f$ is finite at the transition such that the transition is discontinuous.

\subsection{Hybrid model of $k=2$-core and $k=3$-core}
To search for a tricritical point, we define a model where some fraction of $k=k_1$ vertices, $f$, and the remaining fraction as $k=k_2$ vertices. In this model, normalization demands that
\begin{eqnarray}
1&=&\sum_{z=0}^{\infty}P_z(t)\nonumber\\
&=&f\sum_{z=0}^{\infty}P_z(t)+(1-f)\sum_{z=0}^{\infty}P_z(t)\nonumber\\
&=&f(\sum_{z=0}^{k_1-1}P_z(t)+\sum_{z=k_1}^{\infty}P_z(t))+(1-f)(\sum_{z=0}^{k_2-1}P_z(t)+\sum_{z=k_2}^{\infty}P_z(t))
\end{eqnarray}
As before, we will assume that for the $k_1$-core vertices, as long as $z\geq k_1$, their Poissonian structure is retained with some effective connectivity that changes as a function of time.  We will assume this property for the $k_2$-core vertices as well with both types of vertices having the same effective connectivity since they are part of the same graph. Therefore,
\be
1=f(\sum_{z=0}^{\infty}\chi_{k_1}P_z(t)+\sum_{z=k_1}^{\infty}\frac{1}{1-t}e^{-\beta(t)}\frac{\beta(t)^d}{d!})+(1-f)(\sum_{z=0}^{\infty}\chi_{k_2}P_z(t)+\sum_{z=k_2}^{\infty}\frac{1}{1-t}e^{-\beta(t)}\frac{\beta(t)^z}{z!}),
\ee
where $\chi_{k_1}$ applies to the $k_1$-core vertices with $\chi_{k_1}=0$ for $z\geq k_1$ and is unity otherwise, and similarly, for $\chi_{k_2}$ with $k_2$ replacing $k_1$, to arrive at
\be
1=f\overline{\chi_{k_1}}+\frac{1}{1-t}F_{k_1}(\beta(t))+ (1-f)(\overline{\chi_{k_2}}+\frac{1}{1-t}F_{k_2}(\beta(t))).
\ee
Moreover, the average connectivity can be written as:
\be
Z(t)=f(\overline{z\chi_{k_1}}+\frac{\beta(t)}{1-t} F_{k_1-1}(\beta(t)))+(1-f)(\overline{z\chi_{k_2}}+\frac{\beta(t)}{1-t} F_{k_2-1}(\beta(t))).
\ee

We, again, use the rate equation to obtain a relation between the average connectivity and the effective connectivity.  The rate equation for the hybrid model is
\be
\frac{d}{dt}\{(1-t)P_z(t)\}=f\{-\frac{\chi_{k_1}P_z(t)}{\overline{\chi_{k_1}}}+\frac{\overline{z\chi_{k_1}}}{\overline{\chi_{k_1}}}[-\frac{zP_z(t)}{Z(t)}+\frac{(z+1)P_{z+1}(t)}{Z(t)}]\}+(1-f)\{-\frac{\chi_{k_2}P_z(t)}{\overline{\chi_{k_2}}}+\frac{\overline{z\chi_{k_2}}}{\overline{\chi_{k_2}}}[-\frac{zP_z(t)}{Z(t)}+\frac{(z+1)P_{z+1}(t)}{Z(t)}]\}.
\ee
Using the Possionian ansatz, the RHS of the rate equation becomes
\begin{eqnarray}
& &f\frac{\overline{z\chi_{k_1}}}{\overline{\chi_{k_1}}}[-e^{-\beta(t)}\frac{\beta(t)^z}{(z-1)!(1-t)Z(t)}+e^{-\beta(t)}\frac{\beta(t)^{z+1}}{z!(1-t)Z(t)}]\nonumber\\
&+&(1-f)\frac{\overline{z\chi_{k_2}}}{\overline{\chi_{k_2}}}[-e^{-\beta(t)}\frac{\beta(t)^z}{(z-1)!(1-t)Z(t)}+e^{-\beta(t)}\frac{\beta(t)^{z+1}}{z!(1-t)Z(t)}]
,\end{eqnarray}
and the LHS is
\be
\dot{\beta(t)}[e^{-\beta(t)}\frac{\beta(t)^{z-1}}{(z-1)!}-e^{-\beta(t)}\frac{\beta(t)^z}{z!}]
\ee
to arrive at
 \be
\dot{\beta(t)}=-\frac{\beta(t)}{m(t)}(f\frac{\overline{z\chi_{k_1}}}{\overline{\chi_{k_1}}}+(1-f)\frac{\overline{z\chi_{k_2}}}{\overline{\chi_{k_2}}})
,\ee
where $m(t)=(1-t)Z(t)$. It turns out that
\begin{eqnarray} \dot{m(t)}
&=&\frac{d}{dt}\{(1-t)\sum_zzP_z(t)\}\nonumber\\
&=&\sum_zz\frac{d}{dt}\{(1-t)P_z(t)\}\nonumber\\
&=&-f\frac{\overline{z\chi_{k_1}}}{\overline{\chi_{k_1}}}+f\frac{\overline{z\chi_{k_1}}}{\overline{\chi_{k_1}}}[-\frac{\overline{z^2}}{Z(t)}+\frac{\overline{z(z-1)}}{Z(t)}]-(1-f)\frac{\overline{z\chi_{k_2}}}{\overline{\chi_{k_2}}}+(1-f)\frac{\overline{z\chi_{k_2}}}{\overline{\chi_{k_2}}}[-\frac{\overline{z^2}}{Z(t)}+\frac{\overline{z(z-1)}}{Z(t)}]\nonumber\\
&=&-2(f\frac{\overline{z\chi_{k_1}}}{\overline{\chi_{k_1}}}+(1-f)\frac{\overline{z\chi_{k_2}}}{\overline{\chi_{k_2}}})
\label{Eq:dot m(t)}
\end{eqnarray}
such that
\be
\frac{\dot{\beta}(t)}{\beta(t)}=\frac{1}{2}\frac{\dot{m}(t)}{m(t)}
\ee
whose solution is $m(t)=\frac{\beta^2(t)}{c}$ as before.

Now we are ready to extract the criticial behavior for this hybrid model by occupying edges at random with an initial average connectivity $Z(0)=c$ and iterating the culling process until a stable configuration is found at time $t=t_f$. Given the $k$-core constraints, when $t=t_f$, $\overline{\chi_{k_1}}=\overline{\chi_{k_2}}=\overline{z\chi_{k_1}}=\overline{z\chi_{k_2}}=0$ such that the normalization condition becomes
\be
1=f\sum_{z=k_1}^{k_2-1}\frac{1}{1-t_f}e^{-\beta_f}\frac{\beta_f^z}{z!}+\sum_{z=k_2}^\infty\frac{1}{1-t_f}e^{-\beta_f}\frac{\beta_f^z}{z!}\label{eq:Normalization.1}
\ee
with $\beta_f=\beta(t_f)$.
The average connectivity equation becomes
\be
Z(t_f)=f\sum_{z=k_1}^{k_2-1}\frac{z}{1-t_f}e^{-\beta_f}\frac{\beta_f^z}{z!}+\beta_f\frac{1}{1-t_f}F_{k_2-1}(\beta_f).
\ee
For $k_1=2$ and $k_2=3$,
\be
Z(t_f)=\frac{2f}{1-t_f}e^{-\beta_f}\frac{\beta_f^2}{2}+\frac{\beta_f}{1-t_f}(1-e^{-\beta_f}(1+\beta_f)).
\ee
Using Eq. 8, we arrive at
\be
\frac{\beta_f}{c}=fe^{-\beta_f}\beta_f+1-e^{-\beta_f}(1+\beta_f)=F(\beta_f).
\label{Eq:betaequation}
\ee
As before, this self-consistency equation determines $\beta_f$ and then one can use the normalization condition at $t=t_f$ to find the size of the giant component. See Figures 1-3 for a graphical representation of this equation for different values of $f$.

When $f=1$, the model reduces to $k=2$-core, and the transition is continuous. When $f=0$, the model reduces to $k=3$-core, and the transition is discontinuous.  As $f$ is varied between zero and unity, the curvature of $F(\beta_f)$ at $\beta_f=0$ changes from positive to negative such that for some particular value of $f$, the curvature of $F(\beta_f)$ at $\beta_f=0$ vanishes.  In other words, there exists a tricritical point separating the continuous $k=2$-core transition from the discontinuous $k=3$-core transition.  This tricritical point occurs at $f=1/2$. We will first examine the scaling at the tricritical point and then for $f<1/2$ and $f>1/2$.

\begin{figure}[bt]
\begin{center}
\includegraphics[width=10cm]{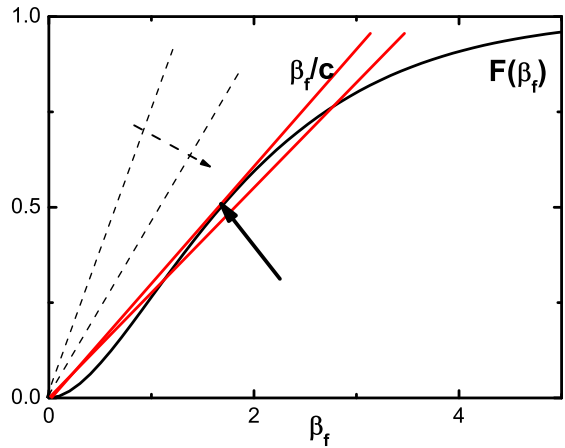}
\caption{For $f<1/2$, the self-consistency equation has two coexisting solutions, $\beta_f=0$ and finte $\beta_f$. The dotted arrow indicates increasing $c$ and the bold arrow indicates the transition point} \label{fig:small.f}
\end{center}
\end{figure}

\begin{figure}[bt]
\begin{center}
\includegraphics[width=10cm]{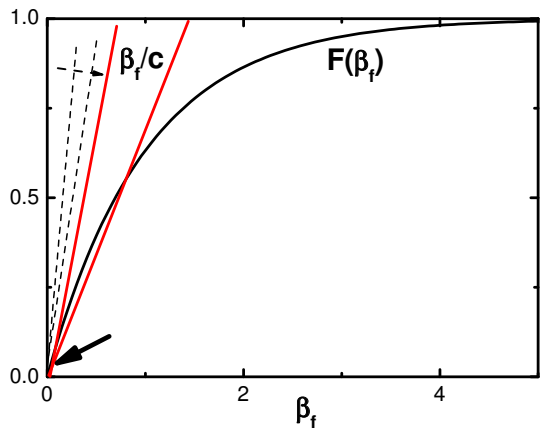}
\caption{For $f>1/2$, the self-consistency equation has one solution for $\beta_f$. The arrows denote the same as in Figure 1.} \label{fig:large.f}
\end{center}
\end{figure}

\subsubsection{The tricritical point: $f=1/2$}

When $f=1/2$, the self-consistency equation (Eq. 23) reads
\be
\frac{\beta_f}{c}=\frac{1}{2}e^{-\beta_f}\beta_f+1-e^{-\beta_f}(1+\beta_f).\label{Eq:beta
equation.f=1/2}
\ee
First we examine the scaling of $\beta_f$ with $c$ near the transition.  The critical value of $c$ indicating the onset of the giant component, $c_G$, is determined by $F'(\beta_f)|_{\beta_f=0}=\frac{1}{c}$ yielding $c_G=2$ at the
tricritical point.  Let $c=2+\epsilon$ with $\epsilon<<1$ is a small number and assuming $\beta_f$ changes continuously (as is indicated graphically) such that $\beta_f<<1$, the self-consistency equation becomes
\be
\frac{\beta_f}{2}(1-\frac{\epsilon}{2}+\mathcal{O}(\epsilon^2))=\frac{\beta_f}{2}-\frac{\beta_f^3}{12}+\mathcal{O}(\beta_f^4)
\ee
such that $\beta_f=\sqrt{3}{\epsilon}^{\frac{1}{2}}$.

\begin{figure}[bt]
\begin{center}
\includegraphics[width=10cm]{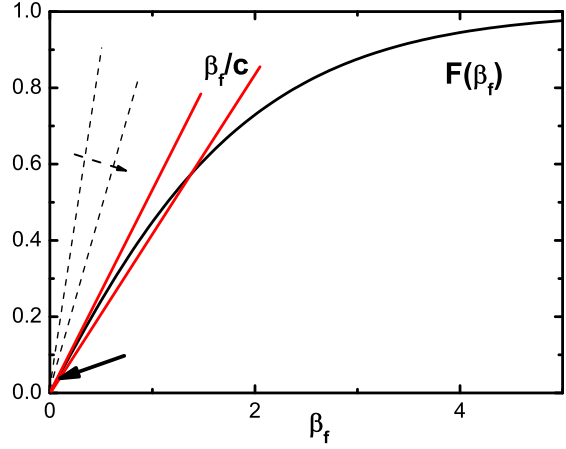}
\caption{For $f=1/2$, the self-consistency equation has one solution for $\beta_f$, but it scales differently with an increase in the initial average connectivity than the $f>1/2$ case. The arrows denotes the same as in Figure 1.} \label{fig:tricritical}
\end{center}
\end{figure}

To find the scaling of the size of the giant component, or $(1-t_f)N$, as a function of the initial average connectivity, $c$, the normalization condition becomes

\begin{eqnarray} 1-t_f
&=&\frac{1}{2}e^{-\beta_f}\frac{\beta_f^2}{2}+(1-\sum_{z=0}^2e^{-\beta_f}\frac{\beta_f^z}{z!})\nonumber\\
&=&1-e^{-\beta_f}-e^{-\beta_f}{\beta_f}-\frac{1}{2}e^{-\beta_f}\frac{\beta_f^2}{2}.
\end{eqnarray}
Expanding in $\beta_f$ yields \be
1-t_f=1-(1-\beta_f+\frac{1}{2}\beta_f^2-\frac{1}{6}\beta_f^3+\mathcal{O}(\beta_f^3))(1+\beta_f+\frac{1}{4}\beta_f^2)
\ee such that $1-t_f=\frac{1}{4}\beta_f^2$ leads to \be
1-t_f=\frac{3}{4}{\epsilon}. \ee This scaling relation is
consistent with Ref.~\cite{Cellai}.

One can also vary $f$ and find the scaling of $1-t_f$ with $f$. From Eq. 24, instead of changing $c$, we
change $f$ from 1/2 to $1/2+\epsilon$ with $0<\epsilon<<1$.  After expanding in $\epsilon$ and $\beta_f$, $\beta_f\propto\epsilon^{1/2}$. Since the scaling of $\beta_f$ as both $c$ and $f$ are increased beyond the transition, the size of the giant component increases in the same way beyond the transition.

\begin{figure}[bt]
\begin{center}
\includegraphics[width=10cm]{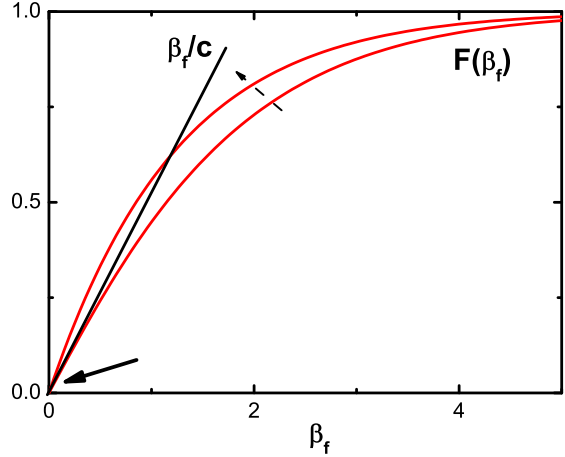}
\caption{Increasing $f$ beyond the tricritical value of $1/2$, one observes that $\beta_f$ changes smoothly. Here, the dotted arrow denotes an increase in $f$.} \label{fig:tricritical.overf}
\end{center}
\end{figure}

\subsubsection{$f<1/2$}
When $f<1/2$, Figure 1 gives us an indication of the scaling behavior near the transition. By increasing the slope of the straight
line, i.e. decreasing $c$, the first solution, $\beta_f=\beta_0>0$,
appears when the line is tangent to $F(\beta_f)$ such that the critical average connectivity is
\be\frac{1}{c_G}=e^{-\beta_0}[(1-f)\beta_0+f]. \label{Eq:differentiation
of beta equation} \ee
Increasing $c_G$ by $\epsilon$ and assuming$\beta_f=\beta_0+A\epsilon^{\lambda}$, where $A$ and $\lambda$ are positive constants, and then the LHS of
Eq. 23 becomes
\be
\frac{\beta_0+A\epsilon^{\lambda}}{c_G+\epsilon}=\frac{1}{c_G}(\beta_0+A\epsilon^{\lambda})(1-\frac{\epsilon}{c_G})=\frac{1}{c_G}(\beta_0-\frac{\beta_0}{c_G}\epsilon+A\epsilon^{\lambda}-\frac{A}{c_G}\epsilon^{\lambda+1}),\label{Eq:beta
equation.f<1/2.LHS}\ee and the RHS becomes \begin{eqnarray} &
&1-e^{-\beta_0-A\epsilon^{\lambda}}[1+(1-f)(\beta_0+A\epsilon^{\lambda})]\nonumber\\
&=&1-e^{-\beta_0}(1-A\epsilon^{\lambda}+\frac{1}{2}A^2\epsilon^{2\lambda}-\frac{1}{6}A^3\epsilon^{3\lambda}+\mathcal{O}(\epsilon^{3\lambda}))[1+(1-f)(\beta_0+A\epsilon^{\lambda})]
.\end{eqnarray}
According to Eq. 30 and Eq. 31, the constant terms and the $\epsilon^{\lambda}$ terms on
 both sides cancel each other, thus, the lowest order of the RHS,
 i.e. $\epsilon^{2\lambda}$, should cancel the $\epsilon$ term on the
 LHS leading to $\lambda=1/2$. To find the scaling of the size of the giant component with the initial average connectivity, $1-t_f\propto\beta_f\propto B+C\epsilon^{1/2}$, where $B$ and $C$ are positive constants.  In other words, the transition is discontinuous.

\subsubsection{$f>1/2$}
Figure 2 qualitatively demonstrates that the onset of a nonzero $\beta_f$ as a
function of $c$ is a continuous transition starting at $\beta_f=0$. The critical value of $c$ is given by
\be
f=\frac{1}{c_G} \label{Eq:f=1/c_G}.\ee
As $c$ is increased by
$\epsilon$ to $c_G+\epsilon$ and assuming $\beta_f$ changes continuously from zero, $\beta_f\sim\epsilon$. Therefore,
\be
1-t_f\sim\epsilon^2
\ee
near the transition. This scaling is to be contrasted with the scaling at the tricritical point where the order parameter exponent is unity.

One can also investigate the scaling of the transition with $f$. As $f$ changes from $f_0$ to
$f_0+\epsilon$, $\beta_f$ changes from $0$ to $A\epsilon^{\lambda}$. Then, the LHS of Eq. 23 becomes
\be\frac{A\epsilon^{\lambda}}{c},\ee
and the RHS becomes
\be
1-(1-A\epsilon^{\lambda}+\frac{1}{2}A^2\epsilon^{2\lambda}-\frac{1}{6}A^3\epsilon^{3\lambda}+\mathcal{O}(\epsilon^{3\lambda}))+(f_0-1+\epsilon)(1-A\epsilon^{\lambda}+\frac{1}{2}A^2\epsilon^{2\lambda}-\frac{1}{6}A^3\epsilon^{3\lambda}+\mathcal{O}(\epsilon^{\lambda}))A\epsilon^{\lambda}
\ee
with \begin{eqnarray}  \textrm{constant term}&=&0;\nonumber\\
{\epsilon}^{\lambda} \textrm{ term}&=&f_0A\epsilon^{\lambda};\nonumber\\
{\epsilon}^{2\lambda} \textrm{
term}&=&(\frac{1}{2}-f_0)A^2\epsilon^{2\lambda};\nonumber\\
{\epsilon}^{\lambda+1} \textrm{
term}&=&A\epsilon^{\lambda+1};\nonumber\\
{\epsilon}^{3\lambda} \textrm{
term}&=&(\frac{f_0}{2}-\frac{1}{3})A^3\epsilon^{3\lambda}.\nonumber\\
\label{Eq:beta equation.f>1/2.RHS}\end{eqnarray}
The $\epsilon^{\lambda}$ terms cancel each other on both sides and
$\epsilon^{\lambda+1}$ and $\epsilon^{2\lambda}$ terms sum to zero leading to $\lambda=1$ and
$A=\frac{1}{f_0-\frac{1}{2}}$. Thus, the scaling of $1-t_f$ with a small change in $f$ is the same as for a small change in $c$. We see that the amplitude diverges when $f_0=1/2$ indicating the vanishing of the $\beta_f^2$ contribution such that the cubic contribution comes into play when $f=1/2$.
\subsubsection{Summary}
Our results for the hybrid $k_1=2$ and $k_2=3$-core on the random graph can be summarized in the phase diagram depicted in Figure 5. For $f>1/2$, the $k_1=2$-core dominates and the transition is continuous with the size of the giant component scaling quadratically with a small increase in the initial average connectivity beyond its critical value. For $f=1/2$ there exists a tricritical point with a new order parameter exponent, and for $f<1/2$ the transition is discontinuous.
\begin{figure}[bt]
\begin{center}
\includegraphics[width=10cm]{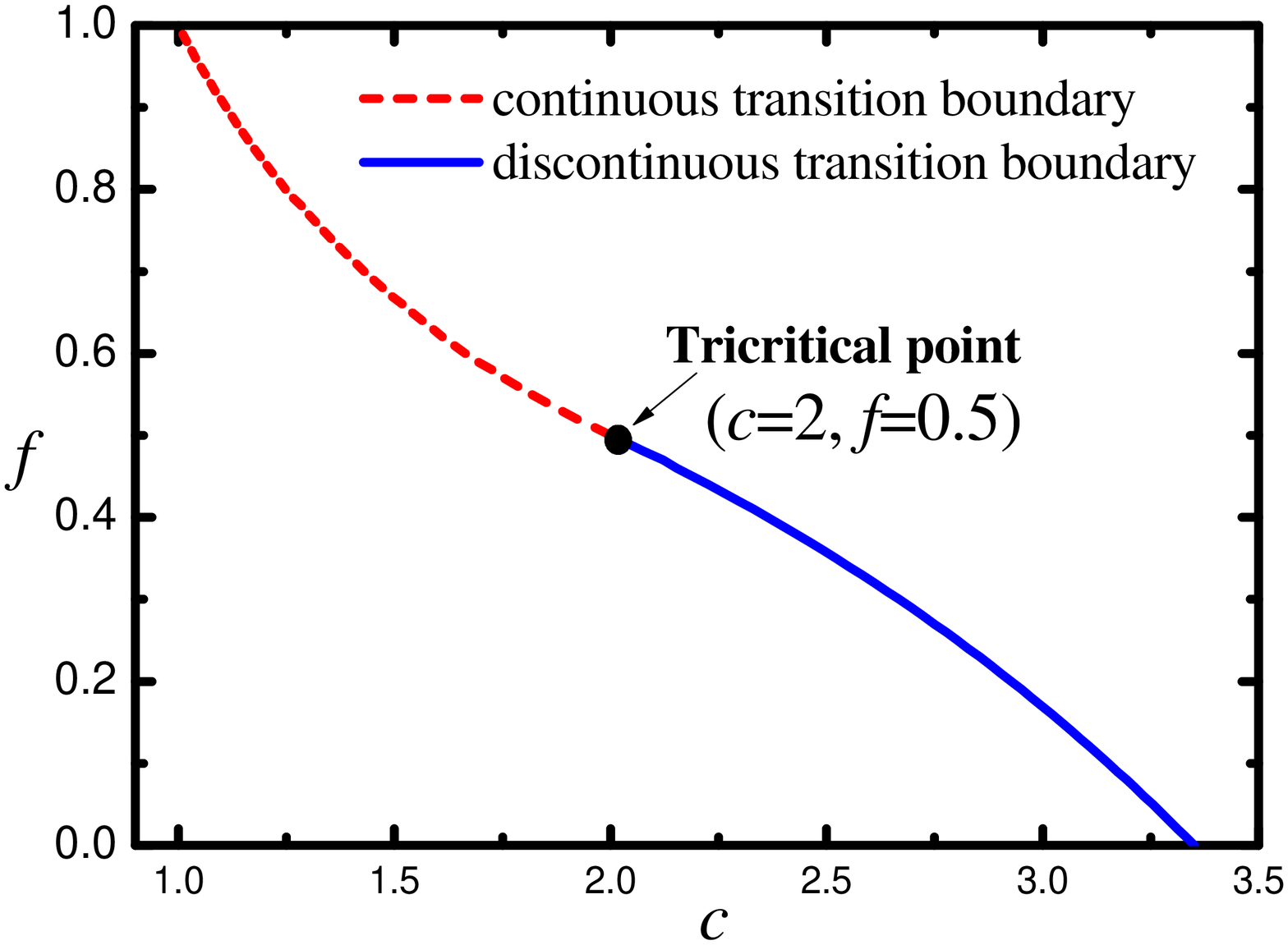}
\caption{The $c,f$ phase diagram. To the left of the boundary there is no giant component and to the right there is.}
\label{fig:phase.diagram}
\end{center}
\end{figure}

\section{Hybrid $k$-core and jamming percolation models in two dimensions}
In order to understand what happens in the hybrid models, we first review two jamming percolation models since they are rather new to the field. These models are vertex models (no edges are explicitly added). Of course, a pure vertex version of $k$-core percolation can also be introduced where an occupied vertex requires at least $k$ occupied vertices to remain occupied. It is this version of $k$-core we will refer to in the following section.
\subsection{Counter-balance model}
While $k\geq 3$-core exhibits a discontinuous transition in mean-field, there is no known two-dimensional example of $k\geq 3$-core exhibiting a discontinuous transition.  In fact, it appears that $k\geq 3$-core on two-dimensional lattices either exhibits an ordinary percolation transition with a shift in the critical occupation probability, or there is no transition until the lattice is fully occupied~\cite{triangular lattice,pc=1}.  For instance, simulations of $k=3$-core percolation on the triangular lattice result in similar ordinary percolation exponents, while for $4\geq k \geq 6$, $p_c=1$. A heuristic argument behind the former result is that finite clusters are allowed for $k=3$ with a fully occupied hexagon being the smallest structure. One can then imagine this object to be smallest object, as opposed to a single vertex, such that path-like spanning structures are formed out of fully occupied hexagons. This procedure is a simple coarse-graining on a microscopic scale and will not effect the macroscopic scales near a continuous phase transition.

Inspired by the jamming transition in two-dimensions where the fraction of particles participating in the jammed structure goes from zero to finite~\cite{jamming1,jamming2}, one can encode various properties of the jammed packings into a percolation model. One of those properties is counter-balancing.  For the force on each particle to be balanced, there must be an occupied particle on either side.  Of course, in two-dimensions, two particles on opposite sides of the particle in question will not suffice since the configuration is not mechanically stable. However, three particles whose centers are 120 degrees with respect to each other as measured from the center particle is a stable configuration.

Counter-balance percolation takes into account the counter-balancing aspect of force-balance~\cite{cb}. As for an example, we begin with a two-dimensional square lattice. Each vertex neighbors
all vertices within a 5x5 square modulo itself. In other words, each vertex has 24 nearest neighbors.  The counter-balancing constraint is the following:
for an initially occupied vertex to remain occupied, there must be at least one occupied neighbor in set A, which
in turn calls for at least one occupied neighbor in set
B, and there must be at least one occupied neighbor in set
C, which in turn calls for at least one occupied neighbor
in set D. The four sets A, B, C, and D, are defined in
Fig. 6. The counter-balance constraint
can be succinctly stated as: (A and B) and (C and D), where
each letter X is short for ``at least one occupied vertex in
set X''. Note that the counter-balance constraint is defined
in such a way such that vertical and/or horizontal lines of
occupied vertices are, by themselves, not stable.
Fig. 7a
demonstrates an allowed configuration and Fig. 7b demonstrates a forbidden configuration.

To enforce the counter-balance constraint,
we initially occupy vertices on the lattice with independent
occupation probabilities $p$, and then repeatedly
remove occupied vertices that violate the counter-balance constraint, until all remaining occupied vertices obey the constraint. Note that $p$ is the occupation density before
culling, and generically differs from the final occupation
density. Moreover, the model is abelian, i.e. the order of the culling does not affect the final configuration.

\begin{figure}[htb]
\begin{center}
\includegraphics[width=5cm]{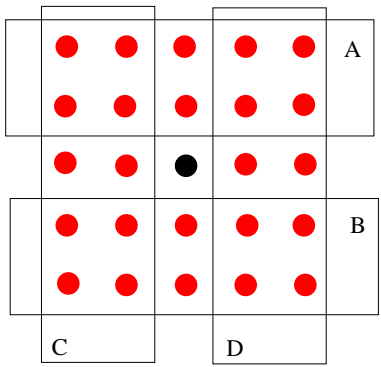}
\caption{Force-balance model on the 2$d$ square lattice
with 24 nearest neighbors.}
\label{fig:z24.definition}
\end{center}
\end{figure}

\begin{figure}[htb]
\begin{center}
\includegraphics[width=2.5cm]{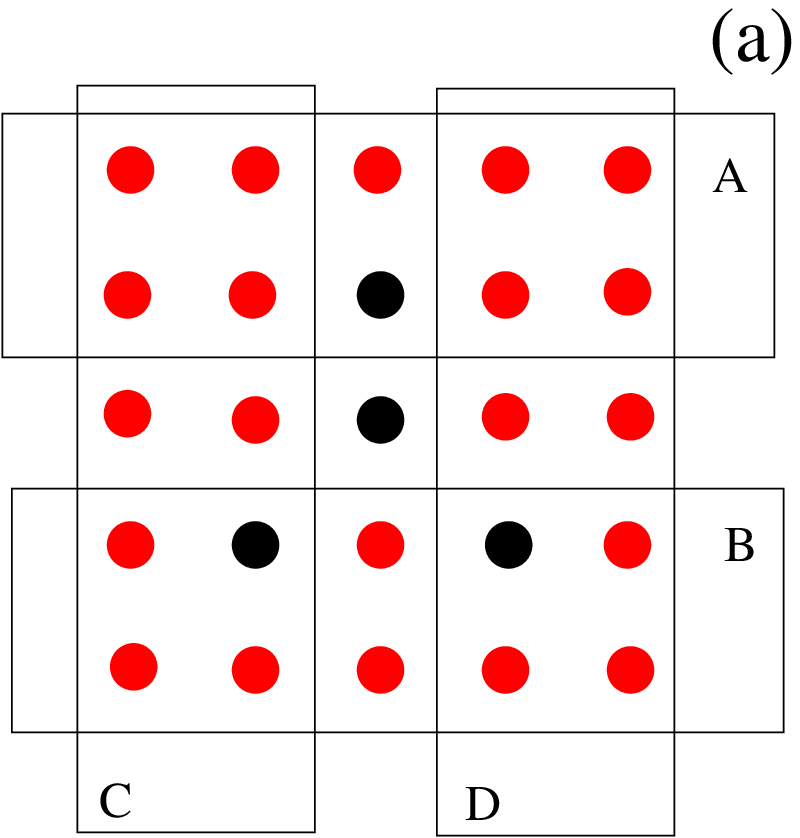}
\includegraphics[width=2.5cm]{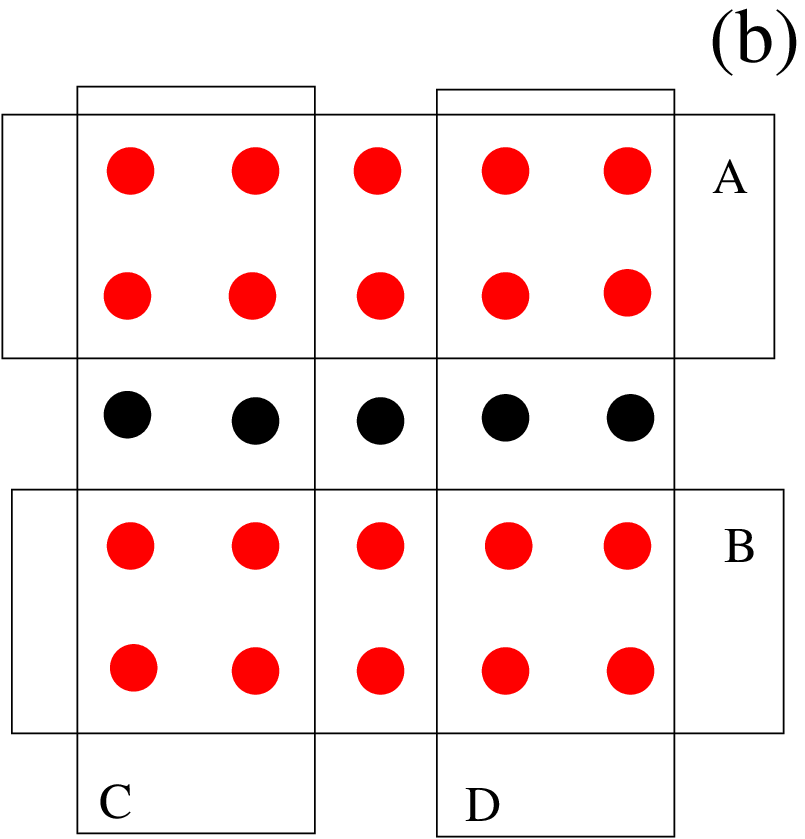}
\caption{Black sites denote occupied sites, and red sites
denote unoccupied sites. (a) Allowed configuration. (b)
Forbidden configuration. While the number of occupied nearest
neighbors is greater than three, the force-balance condition
is violated.}
\label{fig:z24.allow}
\end{center}
\end{figure}

Numerical simulations of the counter-balance model strongly suggest a discontinuous transition with a finite fraction of vertices participating in the spanning cluster {\it at} the transition~\cite{cb}. Of course, numerical simulations can be misleading as has been demonstrated in the past, and so one looks for evidence beyond numerics.  Such evidence is provided for by studying a simpler, but related, jamming percolation model, namely the spiral model. While the spiral model is less physical, one can make several concrete statements about its percolation transition.

\subsection{Spiral model}
The spiral model~\cite{spiralmodel2} is defined as the following.
Again, we begin with a square lattice. The neighbors of each
vertex contain the four nearest neighbors and the four next
nearest neighbors to make a total of eight neighbors. After the
initial random occupation of vertices, for each vertex to remain
occupied there must be at least one occupied neighbor in set A and
at least one occupied neighbor in B, {\it or} there must be at
least one occupied neighbor in set C and one occupied neighbor in
set D. See Fig. 8 to denote the sets.

It turns out that the critical occupation probability for this model is the same as for directed percolation~\cite{dp}, denoted as $p_c^{DP}$. One can see the link with directed percolation when considering only sets A and B or sets C and D.  Each pair of sets is isomorphic to the canonical two-dimensional version of directed percolation with two neighbors above and below each vertex. Therefore, for $p>p_c^{DP}$, there exists a spanning cluster along either diagonal. Since there are two pairs of sets one might argue that the critical occupation probability is lower than that of directed percolation. However, one can show that a certain class of voids (finite ``clusters'' of unoccupied vertices) cause all remaining occupied vertices in the system to become unoccupied as long as $p<p_c^{DP}$. Given the two bounds, $p_c^{SP}=p_c^{DP}$, where $SP$ denotes the spiral model.

\begin{figure}[htb]
\begin{center}
\includegraphics[width=3cm]{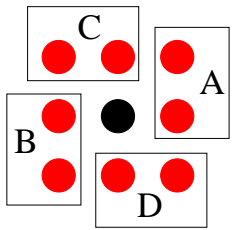}
\caption{The sets in the spiral model.}
\label{fig:spiral}
\end{center}
\end{figure}

While determining $p_c^{SP}$ is a detail, it is an important one for arguing that the percolation transition is discontinuous, which is very different from the continuous transition in directed percolation. To argue for a discontinuity, one can construct a set of spanning structures, which include the origin, and demonstrate that for $p=p_c^{DP}$, the probability of such a set is greater than zero. In other words, the spanning structure is compact at the transition. This argument involves the notion of T-junctions, which characterize one diagonal path being supported by the other diagonal path. For instance, a nonspanning path along one diagonal can survive only if it is sandwiched between two paths of the other diagonal. See Figure 9.

\begin{figure}[htb]
\begin{center}
\includegraphics[width=4cm]{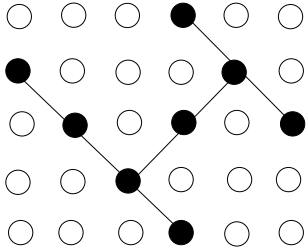}
\caption{A pair of T-junctions in the spiral model.}
\label{fig:tjunction}
\end{center}
\end{figure}

To observe compact spanning structures at the transition, one can
build a spanning structure beginning with a rectangle whose long
side is along the A-B diagonal and includes the origin.  Then, one
considers an infinite sequence of pairs of rectangles of
increasing size emanating outward from the initial rectangle and
intersecting as indicated in Figure 10. If each of the A-B
rectangles contains an A- B spanning path along its length, it is
indeed a spanning structure containing the origin due to the
existence of T-junctions.  From what is known about directed
percolation at the transition, one can show that the probability
for each of the A-B rectangles to contain a A-B spanning path is
indeed greater than zero such that this compact spanning structure
exists {\it at} the transition. Please see
Ref.~\cite{spiralmodel2} for details.

It turns out that one can extend these arguments to jamming
percolation models with more than two pairs of
sets~\cite{momo&jen}. However, these arguments have not yet been
extended to sets with more than two sites. The universality of
directed percolation should allow for such an extension since as
long as the sets are arranged in opposite pairs, the occupation
for each pair of sets is governed by a directed percolation-type
process. It turns out that the rules for the counter-balance model
can be written as a spiral-like model with more than two pairs of
sets with some sets containing more than two vertices. See Figure
11. In addition, the conversion also involves triplets and
quadruplets of sets (as opposed to just pairs) with each set
participates in, for example, a pair-wise interaction as well as a
three-way interaction. These triplets and quadruplets of sets have
not yet been addressed in the context of a spiral-type model but
should not invalidate the overall construction of the above
argument. The fact that one set participates in several
interactions increases the possible configurations, but, again,
should not invalidate the above argument. Therefore, we expect the
percolation transition in the counter-balance model to be
discontinuous with numerical evidence supporting this expectation~\cite{cb}.

\begin{figure}[htb]
\begin{center}
\includegraphics[width=5cm]{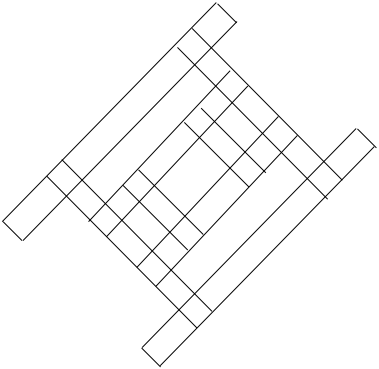}
\caption{Schematic of a spanning structure in the spiral model
where a path spans each rectangle.} \label{fig:spanningspiral}
\end{center}
\end{figure}

\begin{figure}[htb]
\begin{center}
\includegraphics[width=4cm]{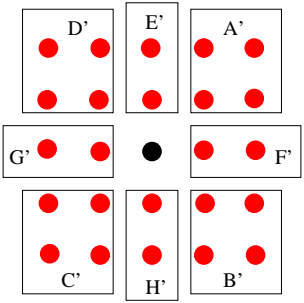}
\caption{The counter-balance rules can be implemented with (A' and C') or (D' and B') or (E' and B' and C') or (F' and C' and D') or (G' and A' and B') or (H' and A' and D') or
(A' and G' and H') or (B' and E' and G') or (C' and E' and F') or (D' and
F' and H') or (E' and F' and G' and H'). }
\label{fig:disjoint}
\end{center}
\end{figure}

\subsection{Tricriticality in the hybrid $k$-core/spiral model?}

For the spiral model, the notion of a T-junction as well as properties of directed percolation are key in establishing a discontinuous transition. What happens when the model is perturbed by having an infinitesimal fraction of vertices with the $k$-core condition replacing the spiral condition?  To answer this question, we must first choose $k$. The value of $k$ is chosen such that $p_c$ for the lattice with every vertex obeying the $k$-core condition is less than $p_c^{DP}$.  For example, $k=3$ satisfies this condition. Note that each vertex has the same eight neighbors as in the spiral model. In addition, when all vertices are $k$-core vertices, we assume that the transition is continuous and in the same universality class as ordinary percolation. This model is similar to the $k=3$-core transition on the triangular lattice.  In both cases, there exist small clusters that survive the culling process such that, heuristically, these clusters link up to form path-like structures as occurs in ordinary percolation.

Given the properties of the all $k$-core transition, as the fraction of spiral model vertices increases from zero, the critical occupation probability increases since the $k$-core condition is less constraining than the spiral model condition.  As long as the critical occupation probability is less than $p_c^{DP}$, the construction invoking a spanning scaffold of T-junctions at $p_c^{DP}$ to demonstrate a discontinuous transition for the all spiral model vertices no longer holds.  The integrity of the T-junctions to arrive a spanning structure is destroyed with finite clusters now allowed even for an infinitesimal fraction of $k$-core vertices. T-junctions are no longer necessary to support a structure. This property is consistent with the violation of the former bound, $p>p_c^{DP}$, since each diagonal is no longer isomorphic to directed percolation independently. The second former bound,
$p<p_c^{DP}$, also breaks down since the growth of voids is now stopped by $k$-core vertices.

The existence of finite clusters for any fraction of $k$-core
vertices is certainly different from the spiral model where no
finite clusters are allowed. Does the existence of finite clusters
imply that the transition is continuous up until all $k$-core
vertices are replaced with spiral model ones? In fact, there's no
direct relationship between the existence of finite clusters and
the continuity of a transition. One can construct a model
occupying vertices at random and independently and then remove all
finite clusters. While this particular constraint is highly
non-local, it preserves the properties of the spanning cluster (of
ordinary percolation) at the transition. Even so, we argue that
the transition is continuous for the following reason. As the
fraction of spiral model vertices increases, their increasing
presence demands the increasing use of T-junctions to support
spiral model vertices where two paths along one diagonal sandwich
and support the path along the second diagonal. However, these two
supporting paths can now end on $k$-core sites as opposed to other
T-junctions to form finite clusters. Two finite clusters whose
interiors each contain spiral model sites can join in such a way
that the removal of one site in the two cluster formation that is
not shared by both clusters before the joining does not induce the
removal of the other cluster. We speculate that this independent
cluster joining property leads to a continuous transition since
the joining of such clusters leads to path-like structures on a
larger scale as clusters are placed ``side-by-side''.  The larger
the clusters, the larger the scale one has to go to ``observe''
the path-like structures. See Figure 12.

One would like to make the speculation that the independent cluster joining property leads to a continuous transition more rigorous. Such speculation may lead to a framework to prove that the transition for similar models, such as $k=3$-core on the triangular lattice, is in the same universality class as ordinary percolation. Currently, there is only numerical evidence for $k=3$-core on the triangular lattice being in the same universality class as ordinary percolation.

\begin{figure}[htb]
\begin{center}
\includegraphics[width=5cm]{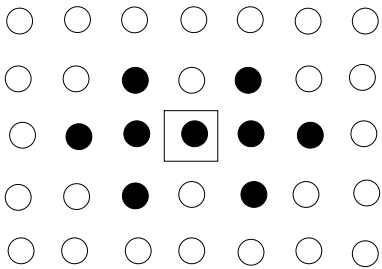}
\caption{An example of two finite clusters for $k=3$-core in the eight neighbor model sharing a boxed occupied vertex.  The removal of any occupied vertex to the right of the boxed occupied vertex does not trigger the removal of any occupied vertex to the left of the boxed occupied vertex.}
\label{fig:finitecluster}
\end{center}
\end{figure}

So, even for an infinitesimal fraction of $k$-core vertices with all other vertices dictated by the spiral model, we speculate that the transition is in the same universality class as ordinary percolation. In other words, only when all vertices are spiral model vertices is there a discontinuous transition. Given this scenario, we speculate that there is no tricritical point in this hybrid of $k=3$-core and the spiral model.

\subsection{Tricriticality in hybrid $k$-core/counter-balance model?}
Since the counter-balance model can be expressed as spiral
model-type constraints, what happens in the hybrid $k$-core/spiral
model may resemble what happens in the hybrid
$k$-core/counter-balance model.  Therefore, we expect that the
transition is continuous as long as $1-f=g<1$, where $g$ is the
fraction of counter-balance vertices, with $k=3$. We now provide
numerical evidence to support this claim.

First, in Figure 13 we plot the differential curve for the probability of spanning, $P_{span}$, as a function of $p$ for $g=0$ and $k=3$ with 24 nearest neighbors on the square lattice. The system length is denoted by $L$ and periodic boundary conditions are implemented.  The position of the peak denotes the critical occupation probability, $p_c$, for each particular system size.  We then perform a scaling collapse using the correlation length exponent of ordinary percolation $\nu=4/3$. See the inset to Figure 13. The scaling collapse is reasonable suggesting that the transition is at least consistent with the ordinary percolation universality class as should be the case for $g<1$ given the discussion in the previous subsection.

\begin{figure}[t]
\begin{center}
\includegraphics[width=9cm]{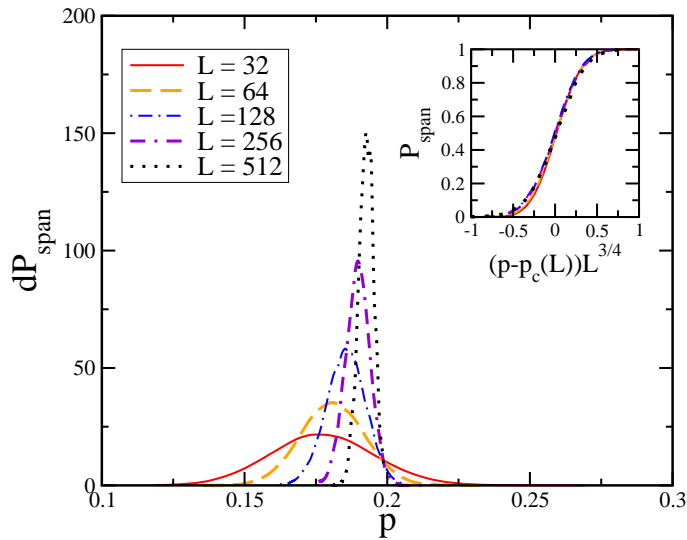}
\caption{Differential for the probability of spanning, $P_{span}$, as a function of occupation probability $p$ for $g=0$. Different system sizes are studied. The inset demonstrates the finite-size scaling for $P_{span}(p)$ using the correlation length exponent for ordinary percolation.}
\end{center}
\end{figure}

Next, we plot average size of any spanning cluster, $<S>$ (the brackets denote the configuration averaging) as a function of the initial occupation probability $p$ for different $g$s.  See Fig. 14. In an infinite system, for $g<1$, we expect
$<S>=A(g)(p-p_c(g))^{5/36}$ for $p>p_c(g)$ and zero otherwise.  In a finite system, this function will be smoothed out.  As expected, we observe that $p_c(g)$ increases with increasing $g$. The curves, even for $g=1$, appear to be sharpening with increasing $g$ as well such that the amplitude $A(g)$ increases with increasing $g$. However, the $<S>$ versus $p$ curves appear qualitatively similar, so how does one differeniate between a continuous transition and a discontinuous one?

We can do so by examining the distribution of the sizes of the spanning cluster as opposed to just looking at the average size. See Figs. 15 and 16. To compare distributions among the different $g$s, we plot the distribution for the same average spanning size, which means $p$ changes from curve to curve. The distribution for $g=0$ has a well-defined peak at small values of $S$, small meaning much less than unity.  As $g$ increases from zero, the distribution for $S$ looks similar with a slight shift in the distribution. It is not until $g\approx$ 0.99 that the position of peak increases such that it is much greater than the fixed average size of 0.05 indicating that in many instances no spanning cluster found and when they are found, the spanning cluster is much larger than the average.  This feature of the distribution, as long as it persists in the infinite system limit, is characteristic of a discontinuous percolation transition.  When the system size is increased for fixed $g$, we observe that the position of the peak shifts to the left indicating that the transition becomes more continuous even for $g=0.99$. Therefore, the numerical data supports our scenario of the absence of a discontinuous transition until $g=1$ and no tricritical point.

\begin{figure}[t]
\begin{center}
\includegraphics[width=9cm]{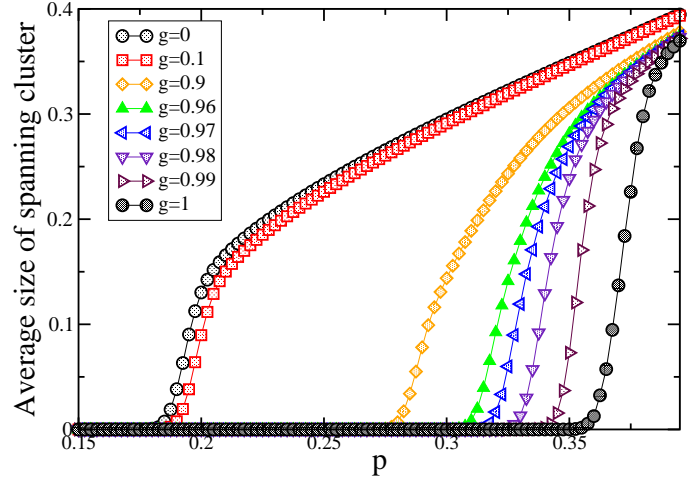}
\caption{Average size of spanning cluster, $<S>$, as a function of $p$ for a mixture of $k=3$-core and counter-balance percolation vertices. For $g=0$, the vertices are all $k$-core.  For $g=1$, the vertices are all counter-balance. Even for $g=0.96$, the curve resembles the $g=0$ (continuous) case. Here, the system length, $L$, is $L=256$. }
\end{center}
\end{figure}

\begin{figure}[b]
\begin{center}
\vspace{0.5cm}
\includegraphics[width=9cm]{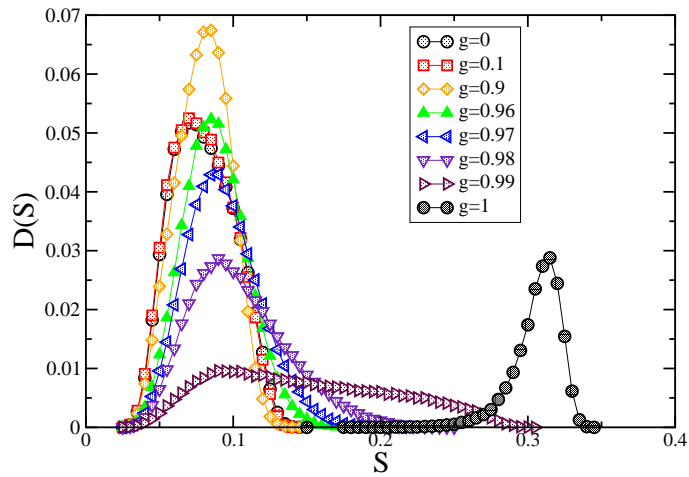}
\caption{Distribution of the size of spanning cluster, $S$, for different $g$s. For $g=1$, the vertices are all counter-balance. Only for $g=0.99$ does the distribution begins to shift to the right significantly, whereas for $g=0.98$ the distribution more closely resembles the $g=0$ case. Here, again, $L=256$. The no spanning cluster contributions are not shown.}
\end{center}
\end{figure}

\begin{figure}[h]
\begin{center}
\includegraphics[width=9cm]{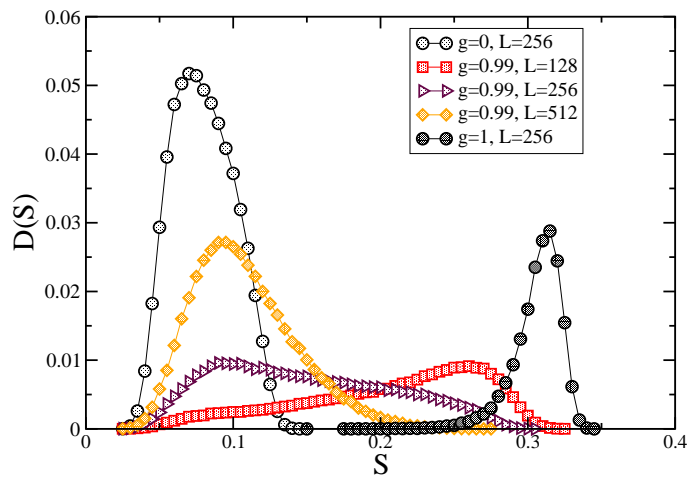}
\caption{Distribution of the size of spanning cluster, $S$, for different $g$s and for different system lengths. Note that as $L$ increases, the curves for $g=0.99$ shift to the left indicating that as $L$ becomes large, the transition approaches the $g=0$ continuous case as opposed to the $g=1$ discontinuous case. }
\end{center}
\end{figure}
\section{Discussion}

There has been a recent proliferation of percolation models going beyond the original model, where edges/vertices are randomly and independently added to a network, to find discontinuous transitions. While much of this proliferation is driven by what is called explosive percolation, there has been, shall we say, quieter progress on what is called jamming percolation models inspired by jamming and glassy systems.  Interestingly, there exists at least one example of a discontinuous percolation transition on random graphs, called $k$-core percolation, which was posed back in 1979~\cite{CLR,Wormald3authors}. However, this transition is not the ordinary type of discontinuous transition lacking any diverging lengthscales.  In fact, the $k\geq 3$-core transition exhibits several diverging lengthscales. In addition, the spiral model, and its related counter-balance model, provide examples of discontinuous percolation transitions in two-dimensions, again, with diverging lengthscales~\cite{spiralmodel2,cb}.

Since both $k\geq 3$-core and the spiral model exhibit discontinuous transitions, it may be useful to investigate what properties they share in order to execute a more efficient search for other discontinuous percolation transitions. Both models share the property of not ``allowing'' finite clusters.  In the spiral model, this property is exact. For the $k$-core model on trees, this property is exact. On random graphs, this property is approximate but becomes an increasingly better approximation in the thermodynamic limit. One may argue that forbidding finite clusters implies a discontinuous transition, at least in low-dimensions. However, this is not the case, since one can construct the usual percolation model on the square lattice and remove any finite clusters using a non-local rule.  One can also do this using a local rule using only one pair of sets from the spiral model.

The nontrivial nesting of the spanning structure in the spiral
model with the two diagonals interdependent on one another tells
us that even if finite clusters are allowed, the formation and
subsequent joining of even finite clusters should be non-local in
the sense the removal of one vertex in a cluster signals the
removal of at least a finite fraction of vertices in its formerly
disjoint neighboring clusters.  The rarefication~\cite{rarefication}
of the lattice or a cluster aggregation model with a non-local
kernel~\cite{clusteraggregation,clusteraggregation2} provide such mechanisms for
nontrivial cluster joining such that the transition is
discontinuous.  On the flip side, we speculate that perhaps the
independent finite cluster joining property, where the removal of
one vertex in one of the two finite clusters (determined prior to joining) that is not shared by
the other does not induce the removal of the other finite cluster.
This property provides for a simple path-like joining of finite
clusters such that an ordinary percolation transition presumably
occurs as is found in $k\geq 3$-core models in finite dimensions.

Given the existence of both continuous and discontinuous percolation transitions, one may form hybrid models of the two types to search for a tricritical point separating the continuous regime from the discontinuous regime. Using a mixture of $k=2$-core and $k=3$-core vertices on random graphs, we find a tricritical point and determine the size of the giant component as a function of the average connectivity. While two previous works have identified this point~\cite{Cellai,Branco}, we use a different dynamical method via a rate equation approach, which could prove to be valuable for investigating other correlated percolation models as the constraints become complex. Of course, our results using the rate equation approach agree with previous results using a ``static'' method.

Moreover, we investigate the possibility of a tricritical point in
two-dimensions. We do so with a mixture of $k=3$-core vertices and
counter-balance vertices where the full $k=3$-core model exhibits
a continuous transition (which differs from mean-field) and the
full counter-balance model exhibits a discontinuous phase
transition. We argue that there is no tricritical point in this
mixed model since the discontinuous transition occurs only in the
full counter-balance model. However, we expect interesting
crossover behavior between to the two types of transitions that
should be explored.  This result is the first we know of with a
line of continuous transitions ending at a discontinuous
transition, which is to be compared with the water phase diagram
where there is a line of discontinuous transitions ending at a
continuous one. In addition, this result is to be contrasted with
the work of Cellai and collaborators studying a mixture of $k=2$
and $k=3$-core vertices on the square lattice~\cite{Cellai}. For
the full $k=2$-core model on the square lattice, the transition is
continuous, while for the full $k=3$-core model, $p_c=1$ with
interesting crossover behavior between the two cases. This result
is also to be contrasted with the tricriticality obtained in a
diluted $Q$-state Potts model on a triangular
lattice~\cite{tricriticalPotts}.

Finally, what about ties between mixed correlated percolation
models and physical systems? The introduction of hetereogeneities
into the Frederickson-Anderson model for glassy systems maps to
the mixed $k$-core model~\cite{Sellitto}.  Moreover, there is a
recent experiment~\cite{jammingbyshear} where a two-dimensional
packing of frictional disks is sheared at a packing fraction just
below the shear-free jamming transition to induce jamming. At
small applied shear stress, a subset of the contact network of the
jammed states exhibits spanning structures along one direction
only, while at larger applied shear stress, the spanning
structures (of a subset of the contact network) percolate in both
directions.  One can perhaps model the connectivity of the
spanning structures as the applied shear stress is varied by
taking a mixture of the usual counter-balance model and an
anisotropic version of the counter-balance model where only sets A
and B are considered. More specifically, as the applied shear
stress is increased the ratio of counter-balance vertices to
anisotropic counter-balance vertices increases. However, since
only a subset of the contact network is used in obtaining this
result, some finite structures may be allowed.

JMS acknowledges support from NSF-DMR-CAREER Award 0645373.

\end{document}